\documentclass[12pt]{article}
\pdfoutput=1
\usepackage{subfigure}
\usepackage{amssymb,amsmath}
\usepackage{graphicx}
\usepackage{color}
\usepackage{cancel}
\usepackage[colorlinks=true
,urlcolor=blue
,citecolor=blue
,linkcolor=blue
,pagecolor=blue
,linktocpage=true
,pdfproducer=medialab
]{hyperref}
\usepackage[a4paper,width=15.2cm]{geometry}

\usepackage{ulem}
\usepackage{placeins}

\makeatletter \renewcommand{\@dotsep}{10000} \makeatother
\def\be{\begin{equation}}
\def\ee{\end{equation}}
\def\bea{\begin{eqnarray}}
\def\eea{\end{eqnarray}}
\def\bi{\begin{itemize}}
\def\ei{\end{itemize}}

\def\tst{\tilde t}
\def\ttau{\tilde \tau}

\newcommand\prd[3]{{\it Phys.\ Rev.\ }{\bf D #1} (#2) #3}
\newcommand\prl[3]{{\it Phys.\ Rev.\ Lett.\ }{\bf #1} (#2) #3}

\newcommand\jhep[3]{{\it J. High Energy Phys.\ }{\bf #1} (#2) #3}

\def\tst{\tilde t}
\def\ttau{\tilde \tau}

\newcommand{\beq}{\begin{equation}}
\newcommand{\eeq}{\end{equation}}

\begin{document}
\date{\today}

\begin{center}
{\Large\bf 
NLSP Gluino and NLSP Stop Scenarios \\
\vspace{0.2cm}
from $b$-$\tau$ Yukawa Unification
} \vspace{1cm}
\end{center}

\begin{center}

{\Large
Shabbar Raza$^{a,b,}$\footnote{
Email: shabbar@itp.ac.cn},
Qaisar Shafi $^{b,}$\footnote{
Email: shafi@bartol.udel.edu.
}
and
Cem Salih \"{U}n $^{b,c,}$\footnote{
Email: cemsalihun@uludag.edu.tr}
}

\vspace{0.75cm}

{\it $^a$
State Key Laboratory of Theoretical Physics and Kavli Institute for Theoretical Physics China (KITPC),
Institute of Theoretical Physics, Chinese Academy of Sciences, Beijing 100190, P. R. China
}\\
{\it  $^b$
Bartol Research Institute, Department of Physics and Astronomy,\\
University of Delaware, Newark, DE 19716, USA 
}

{\it  $^c$
Department of Physics, Uluda\~{g} University, TR16059, Bursa, Turkey}

\vspace{1.5cm}
\section*{Abstract}
\end{center}

We present a study of $b$-$\tau$ Yukawa unified supersymmetric $SU(4)_c \times SU(2)_L \times SU(2)_R$ model (with $\mu > 0$), which predicts the existence of gluino - neutralino and stop - neutralino coannihilation scenarios compatible with the desired relic LSP neutralino dark matter abundance and other collider constraints. The NLSP gluino or NLSP stop masses vary between 400 GeV to $\sim$ 1 TeV. The NLSP gluinos will be accessible at the 14 TeV LHC, while we hope that the NSLP stop solutions will be probed in future LHC searches. We also identify regions of the parameter space in which the gluino and the lighter stop are closely degenerate in mass, interchangeably playing the role of NLSP and NNLSP.

We also update a previous study of $t-b-\tau$ Yukawa unification and show that NLSP gluino of mass $\sim 1$ TeV, with a mass difference between the gluino and neutralino of less than 80 GeV, can be realized consistent with the current collider and astrophysical constraints. We present benchmark points for $b-\tau$ and $t-b-\tau$ Yukawa unification that can be accessible at the LHC.

\newpage

\renewcommand{\thefootnote}{\arabic{footnote}}
\setcounter{footnote}{0}



\section{Introduction}
\label{sec:intro}

The discovery of the Higgs boson at the Large Hadron Collider (LHC) \cite{ATLAS, CMS} is a big boost for the Standard Model (SM). Supersymmetry (SUSY) is arguably the prime candidate for beyond the SM physics and the minimal supersymmetric extension of the SM (MSSM) leads in natural way to the gauge coupling unification and provides a solution to the gauge hierarchy problem. In addition, with the assumption of $R-$parity conservation, MSSM also provides a plausible candidate particle for dark matter, namely the lightest supersymmetric particle (LSP). Besides gauge coupling unification, models such as SUSY $SO(10)$ and 
SUSY $SU(4)_c\times SU(2)_{L}\times SU(2)_{R}$ (4-2-2) also suggest $t-b-\tau$ Yukawa Unification (YU)  \cite{big-422,bigger-422,Baer:2008jn,Gogoladze:2009ug,Baer:2009ff,Gogoladze:2009bn,Gogoladze:2010fu}. The 4-2-2 structure allows us to consider non-universal gaugino masses with
\begin{equation}
M_1=\frac{3}{5}M_2+\frac{2}{5}M_3,
\end{equation}
where $M_{1}$, $M_{2}$ and $M_{3}$ are the soft supersymmetry breaking (SSB) mass terms respectively for $U(1)_{Y}$, $SU(2)_{L}$ and $SU(3)_{c}$ gauginos. 

\noindent Supersymmteric $4-2-2$ offers a rich phenomenology, which can be examined in particular at the LHC. As far as we know, it is the only model which requires 
NLSP gluino to bring the relic abundance of LSP neutralino within the observed range of dark matter density in the presence of $t-b-\tau$ YU \cite{Gogoladze:2009ug, Gogoladze:2009bn}. It was also shown that $t-b-\tau$ YU in $4-2-2$ with the same sign SSB gaugino mass terms is compatible with neutralino dark matter through gluino coannihilation channel \cite{Gogoladze:2009ug,Gogoladze:2009bn,Profumo:2004wk,Ajaib1}. Considering opposite sign gauginos with $\mu <0,~M_{2} < 0$ and $M_{3} > 0$ (where $\mu$ is the bilinear Higgs mixing term) in \cite{Gogoladze:2010fu}, $t-b-\tau$ YU consistent with known experimental constraints was achieved in $4-2-2$ for $m_{16} \gtrsim 300$ GeV, as opposed to $m_{16} \gtrsim 8$ TeV for the case of same sign gaugino masses. Here $m_{16}$ denotes the common soft SUSY breaking scalar mass at $M_{GUT}$.

We show in this paper that relaxing $t-b-\tau$ YU to $b-\tau$ YU yields NLSP stop solutions in addition to NLSP gluino. We also find that the NLSP stop is nearly degenerate with the LSP neutralino, and hence the decay $\tilde{t}_{1} \rightarrow c\tilde{\chi}_{1}^{0} $ is the only channel kinematically allowed. 
The ATLAS collaboration has recently searched for such decays for the first time, and the results have excluded NLSP stop up to 270 GeV for LSP neutralino 
with mass of about 200 GeV \cite{ATLAS_NLSPstop}. The CMS collaboration
has ruled out NLSP stop of mass $\lesssim$ 250 GeV, if the mass difference with LSP neutralino is less than 10 GeV \cite{CMS_NLSPstop}. We obtain relatively heavy ($600 \lesssim m_{\tilde{t}_{1}} \lesssim 900$ GeV) NSLP stop solutions, and we hope that future searches will be able to test our results. We also identify some solutions for which the mass difference between the NLSP and NNLSP is small. We find such solutions in both the NLSP stop and NLSP gluino scenarios. 

We also devote a section to $t-b-\tau$ YU in $4-2-2$ to update the results and compare with $b-\tau$ YU. We analyze the data that we obtain in the light of the current experimental constraints including the discovery of the Higgs boson of mass close to 125 GeV, flavor physics and WMAP9. The LHCb collaboration has recently discovered $B_s\rightarrow \mu^{+}\mu^{-}$ with the branching fraction $BF(B_s\rightarrow \mu^{+}\mu^{-})=3.2_{-1.2}^{+1.5}\times 10^{-9}$ \cite{Aaij:2012nna} that is consistent with the SM prediction of $(3.2\pm0.2)\times 10^{-9}$ \cite{Buchalla:1995vs}. In MSSM, this flavor changing decay receives contributions from the exchange of the pseudoscalar Higgs boson $A$ \cite{Choudhury:1998ze},
which is proportional to $(\tan\beta)^6/m_{A}^4$. Since $t-b-\tau$ YU requires large $\tan\beta ~ (\gtrsim 40)$,
it is important to see the impact of $B_s\rightarrow \mu^{+}\mu^{-}$ discovery on 4-2-2 parameter space in the presence of $t-b-\tau$ YU. 
 
The fundamental parameters of the model include

\begin{equation}
m_{16},M_2,M_3,A_0/m_{16},m_{H_d},m_{H_u},\tan\beta,
\label{params}
\end{equation} 
where $m_{16}$ is the universal SSB mass term for sfermions, and $M_{2}$, $M_{3}$ are the SSB gaugino mass terms for $SU(2)_{L}$ and $SU(3)_{c}$ respectively. $A_{0}$ is the universal SSB trilinear interaction coupling, $m_{H_{d}}$ and $m_{H_{u}}$ are SSB mass terms respectively for the up and down type Higgs scalars of the MSSM, and $\tan\beta$ is the ratio of the MSSM Higgs vacuum expectation values (VEVs).

The outline for the rest of the paper is as follows. In section \ref{sec:scan} we summarize the scanning procedure and the experimental constraints applied in our analysis. We present our findings for $b-\tau$ and $t-b-\tau$ YU in section \ref{sec:results}, and we also provide a table with five benchmark points 
that illustrate our results. Our conclusion is summarized in section \ref{sec:conclude}.

\section{Scanning Procedure and Phenomenological Constraints}
\label{sec:scan}
We employ the ISAJET~7.84 package~\cite{ISAJET} 
 to perform random scans over the parameter space 
 given below. 
In this package, the weak scale values of gauge and third 
 generation Yukawa couplings are evolved to 
 $M_{\rm GUT}$ via the MSSM renormalization group equations (RGEs)
 in the $\overline{DR}$ regularization scheme.
We do not strictly enforce the unification condition
 $g_3=g_1=g_2$ at $M_{\rm GUT}$, since a few percent deviation
 from unification can be assigned to unknown GUT-scale threshold
 corrections~\cite{Hisano:1992jj}.
With the boundary conditions given at $M_{\rm GUT}$, 
 all the SSB parameters, along with the gauge and Yukawa couplings, 
 are evolved back to the weak scale $M_{\rm Z}$.

In evaluating Yukawa couplings the SUSY threshold 
 corrections~\cite{Pierce:1996zz} are taken into account 
 at the common scale $M_{\rm SUSY}= \sqrt{m_{\tst_L}m_{\tst_R}}$. 
The entire parameter set is iteratively run between 
 $M_{\rm Z}$ and $M_{\rm GUT}$ using the full 2-loop RGEs
 until a stable solution is obtained.
To better account for leading-log corrections, one-loop step-beta
 functions are adopted for gauge and Yukawa couplings, and
 the SSB parameters $m_i$ are extracted from RGEs at appropriate scales
 $m_i=m_i(m_i)$.
The RGE-improved 1-loop effective potential is minimized
 at an optimized scale $M_{\rm SUSY}$, which effectively
 accounts for the leading 2-loop corrections.
Full 1-loop radiative corrections are incorporated
 for all sparticle masses.

The requirement of radiative electroweak symmetry breaking
 (REWSB)~\cite{Ibanez:1982fr} puts an important theoretical
 constraint on the parameter space.
Another important constraint comes from limits on the cosmological
 abundance of stable charged particles~\cite{Olive}.
This excludes regions in the parameter space where charged
 SUSY particles, such as $\ttau_1$ or $\tst_1$,
 become the LSP.
We accept only those solutions for which one of the neutralinos
 is the LSP and saturates the dark matter relic abundance bound
 observed by WMAP9.

We have performed random scans 
 for the following parameter range:

\begin{eqnarray}
\label{parameterRange}
0 \leq  m_{16}  \leq 20~ {\rm TeV} \nonumber \\
0 \leq  M_{2}  \leq 5~ {\rm TeV} \nonumber \\ 
0 \leq  M_{3}  \leq 5~ {\rm TeV} \nonumber \\
-3 \leq A_{0}/m_{16} \leq 3 \\
2 \leq \tan\beta \leq 60 \nonumber \\
 0 \leq  m_{H_{u}}  \leq 20~ {\rm TeV} \nonumber \\
  0 \leq  m_{H_{d}}  \leq 20~ {\rm TeV} \nonumber
\end{eqnarray}
%
with  $\mu > 0$ and  $m_t = 173.3\, {\rm GeV}$  \cite{:2009ec}.
Note that our results are not too sensitive to one 
 or two sigma variation in the value of $m_t$  \cite{bartol2}.
We use $m_b^{\overline{DR}}(M_{\rm Z})=2.83$ GeV 
 which is hard-coded into ISAJET.

In scanning the parameter space, we employ the Metropolis-Hastings
 algorithm as described in \cite{Belanger:2009ti}. 
The data points collected all satisfy the requirement of REWSB, 
 with the neutralino in each case being the LSP. 
After collecting the data, we impose the mass bounds on 
 all the particles \cite{Olive} and 
 use the IsaTools package~\cite{bsg, bmm} and Ref.~\cite{mamoudi}
 to implement the following phenomenological constraints: 
\begin{eqnarray} 
m_h  = 123-127~{\rm GeV}~~&\cite{ATLAS, CMS}& 
\\
0.8\times 10^{-9} \leq{\rm BR}(B_s \rightarrow \mu^+ \mu^-) 
  \leq 6.2 \times10^{-9} \;(2\sigma)~~&\cite{Aaij:2012nna}& 
\\ 
2.99 \times 10^{-4} \leq 
  {\rm BR}(b \rightarrow s \gamma) 
  \leq 3.87 \times 10^{-4} \; (2\sigma)~~&\cite{Amhis:2012bh}&  
\\
0.15 \leq \frac{
 {\rm BR}(B_u\rightarrow\tau \nu_{\tau})_{\rm MSSM}}
 {{\rm BR}(B_u\rightarrow \tau \nu_{\tau})_{\rm SM}}
        \leq 2.41 \; (3\sigma)~~&\cite{Asner:2010qj}&  
\\
 0.0913 \leq \Omega_{\rm CDM}h^2 (\rm WMAP9) \leq 0.1363   \; (5\sigma)~~&\cite{WMAP9}&
\end{eqnarray} 
As far as the muon anomalous magnetic moment $a_{\mu}$ is concerned, we require that the benchmark
points are at least as consistent with the data as the Standard Model

\section{Results}
{\subsection{NLSP gluino and NLSP stop from $b$-$\tau$ YU}}
\label{sec:results}
\begin{figure}[htp!]
\subfigure{\includegraphics[scale=1]{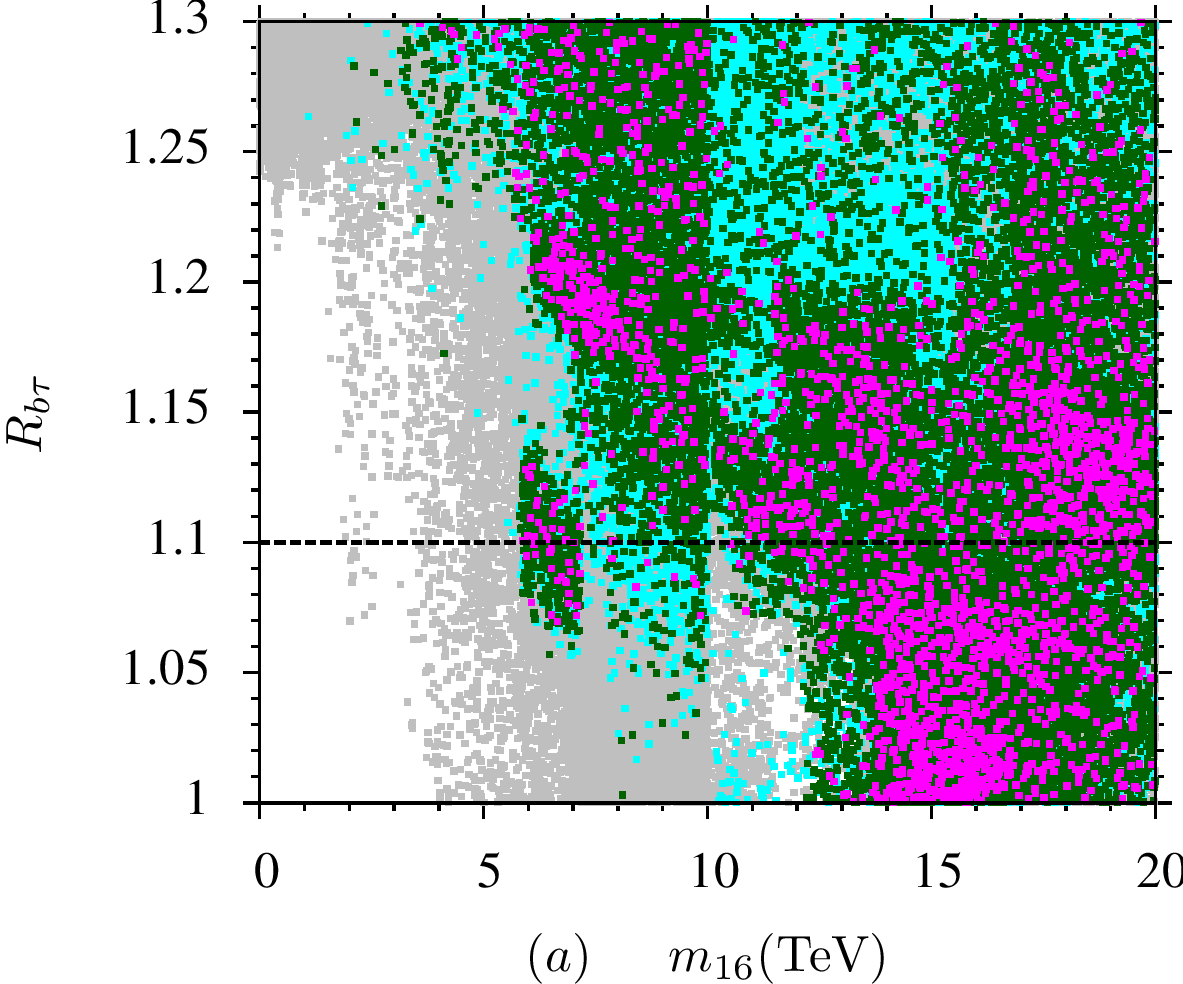}}
\subfigure{\includegraphics[scale=1]{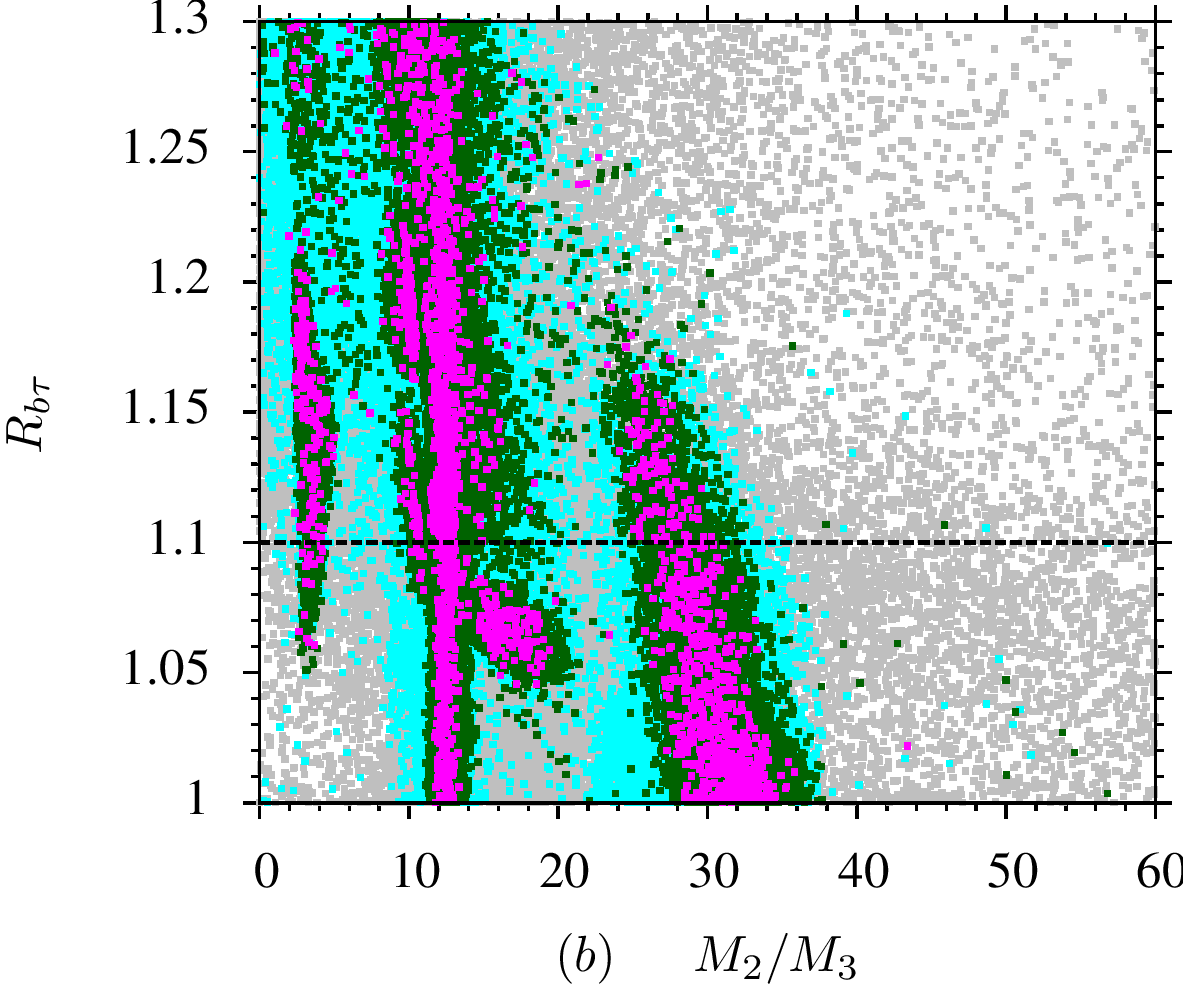}}\\
\subfigure{\includegraphics[scale=1]{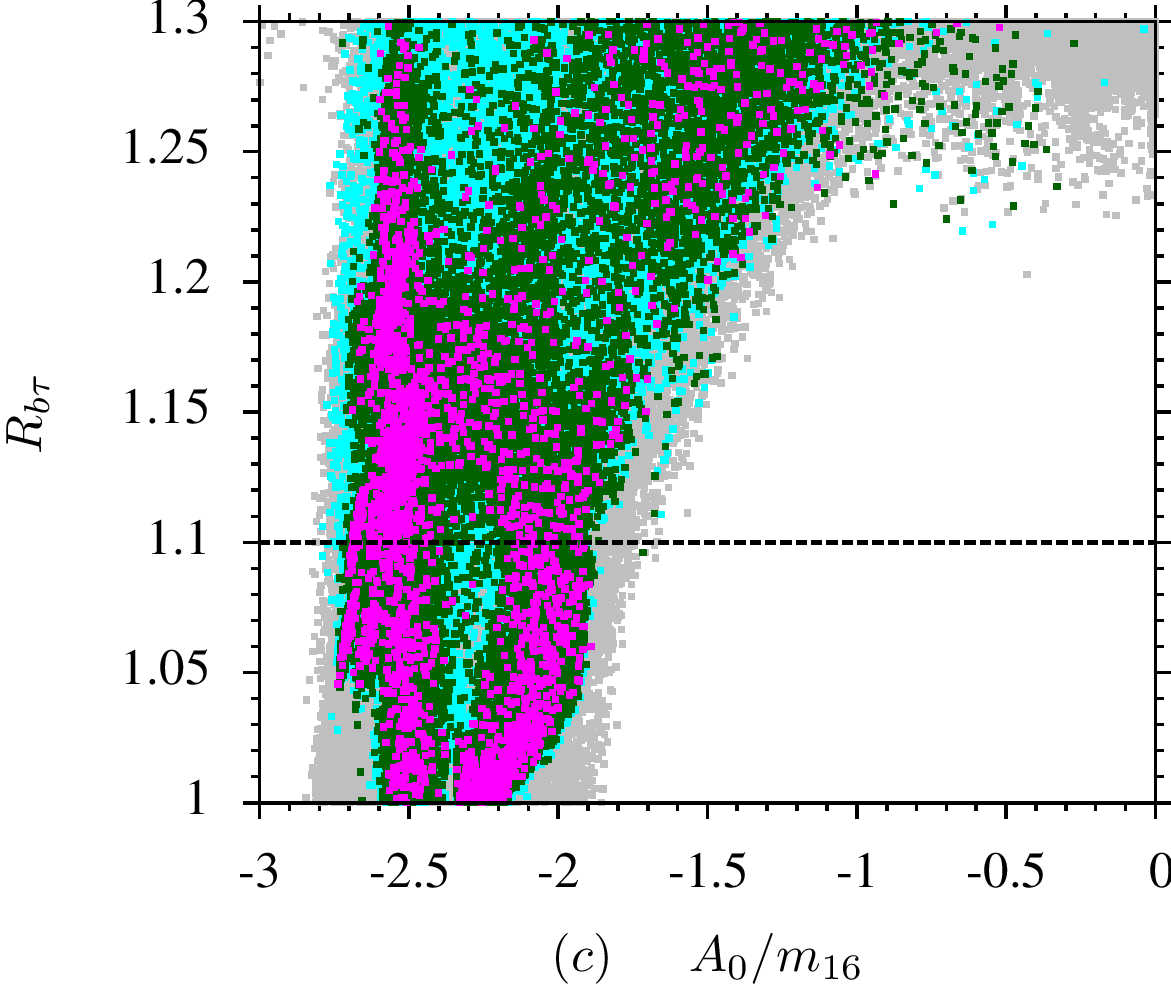}}
\subfigure{\includegraphics[scale=1]{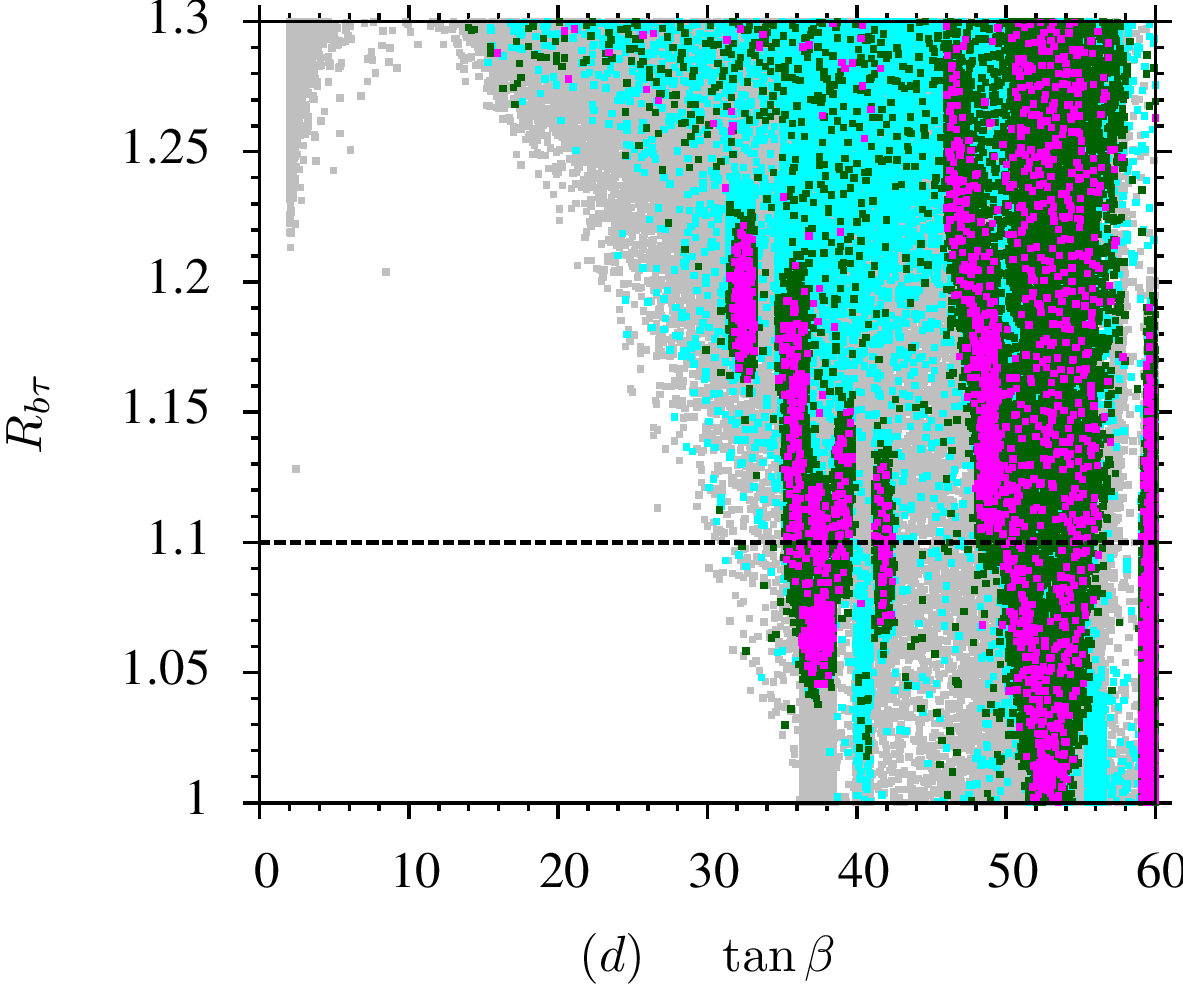}}\\
\subfigure{\includegraphics[scale=1]{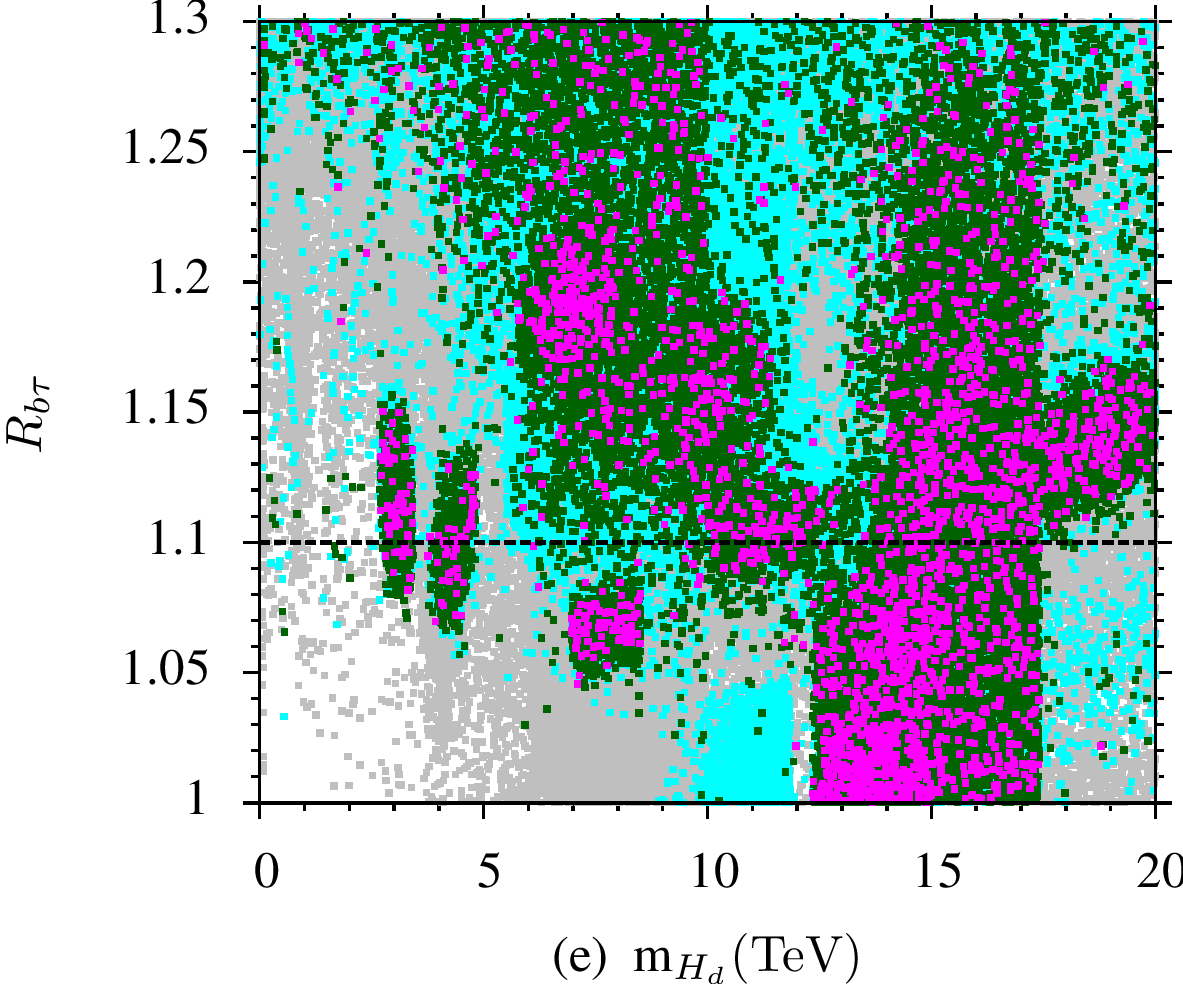}}
\subfigure{\includegraphics[scale=1]{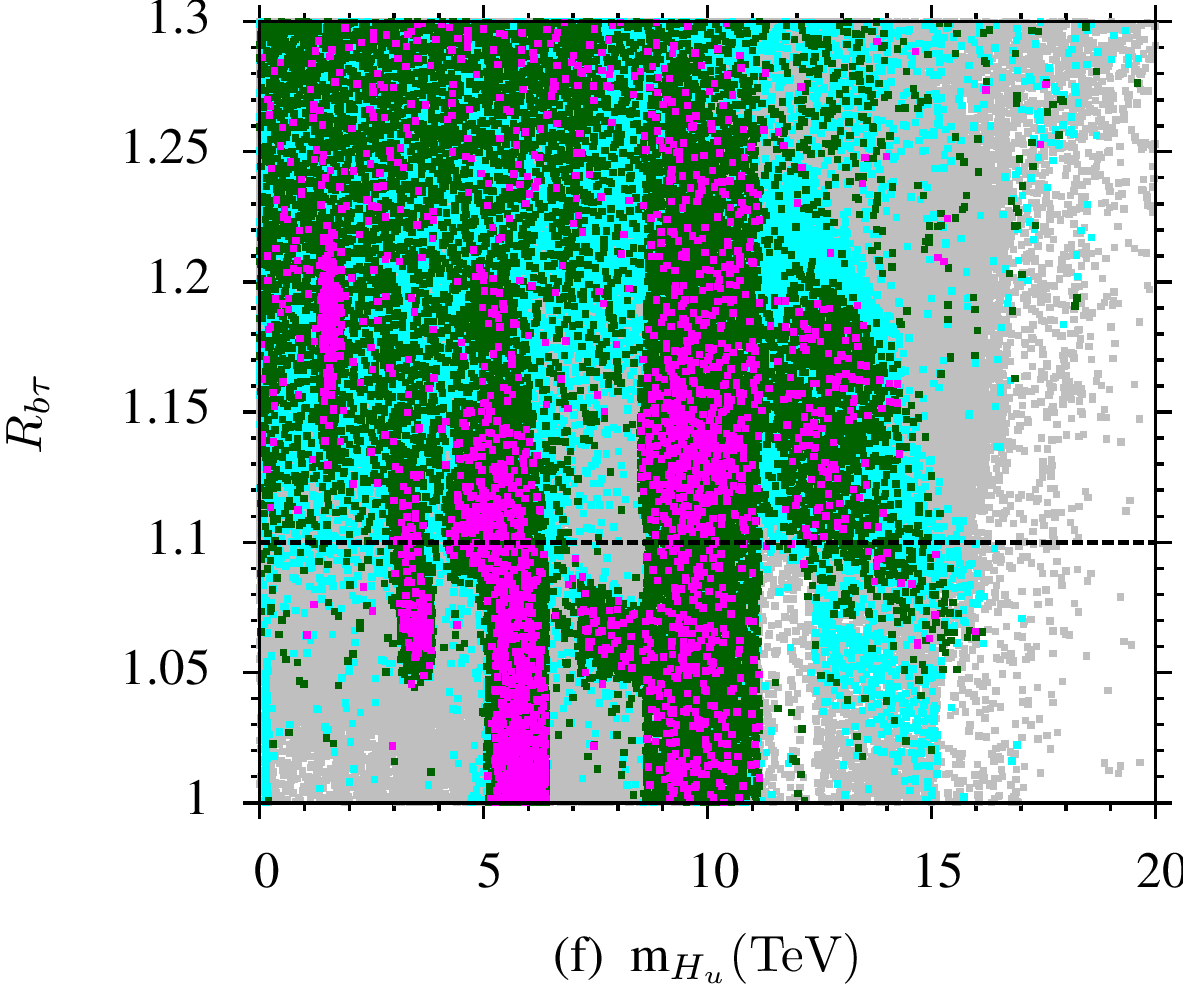}}
\caption{Plots in $R_{b\tau}-m_{16}$, $R_{b\tau}-M_{2}/M_{3}$, $R_{b\tau}-A_{0}/m_{16}$, $R_{b\tau}-\tan\beta$, $R_{b\tau}-m_{H_{d}}$, $R_{b\tau}-m_{H_{u}}$ planes. Grey points are consistent with REWSB and LSP neutralino. Aqua points satisfy mass bounds including bounds on Higgs mass and B-physics constraints. Green points belong to a subset of aqua and represent solutions with $\Omega h^{2} \leq 1$. Magenta points form a subset of green points and satisfy WMAP9 bound on relic abundance of LSP neutralino within $5\sigma$.}
\label{funda}
\end{figure}

We quantify $b-\tau$ YU via the $R_{b\tau}$ parameter defined as \cite{Baer:2008jn}: 
\begin{equation}
R_{b\tau}=\frac{ {\rm max}(y_b,y_{\tau})} { {\rm min} (y_b,y_{\tau})},
\label{eq:R}
\end{equation}
where $y_{b}$ and $y_{\tau}$ are Yukawa couplings at the GUT scale. $R_{b\tau}$ close to uniy denotes acceptable $b-\tau$ YU. In Figure \ref{funda}, we plot $R_{b\tau}$ versus the fundamental parameters given in Eq. \ref{params}. Grey points are consistent with REWSB and LSP neutralino. Aqua points satisfy the various mass bounds which include the Higgs boson, as well as B-physics constraints. Green points belong to a subset of aqua and represent solutions with $\Omega h^{2} \leq 1$. Magenta points form a subset of green points and satisfy the WMAP9 bound within $5\sigma$ on the relic abundance of LSP neutralino . 

In the $R_{b\tau}-m_{16}$ panel, we see that essentially perfect $b-\tau$ YU can be realized for $m_{16} \gtrsim 8$ TeV, while $10\%$ or better $b-\tau$ YU requires $m_{16} \gtrsim 5$ TeV. We see from the $R_{b\tau}-M_{2}/M_{3}$ plot that we can have solutions with perfect $b-\tau$ YU for $M_{2}/M_{3} \gtrsim 10$ in case of green or magenta points, while $b-\tau$ YU within $5-10\%$  is possible when $M_{2}/M_{3} \sim 2$. The trilinear scalar coupling is found from the $R_{b\tau}-A_{0}/m_{16}$ panel to be in the range $-2.7 \lesssim A_{0}/m_{16} \lesssim -2.2 $. The parameter $\tan\beta$ satisfies $30 \lesssim \tan\beta \lesssim 60$. The last two panels of Figure \ref{funda} show the mass range of the MSSM Higgs fields. 


\begin{figure}[h!]
\subfigure{\includegraphics[scale=1]{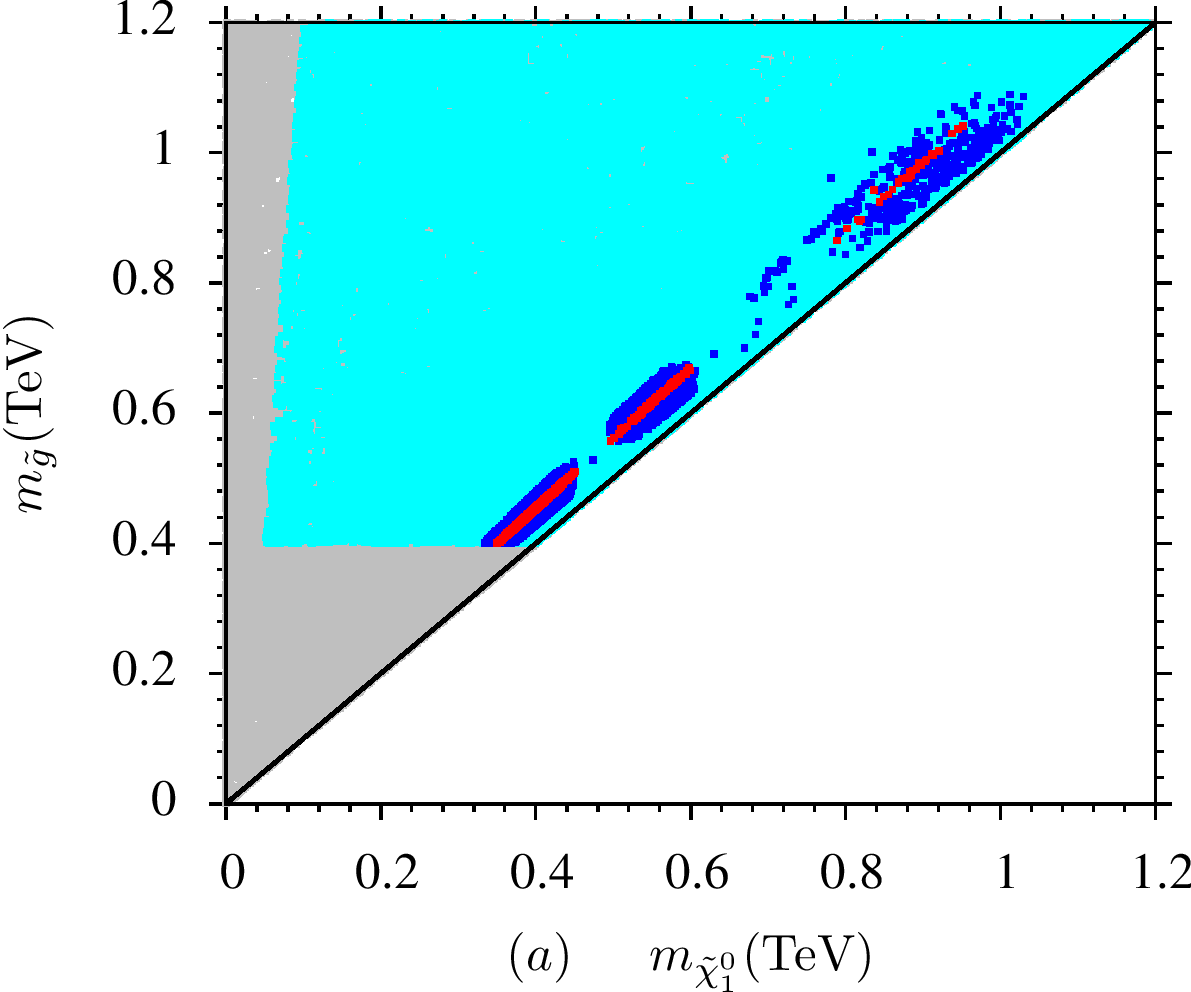}}
\subfigure{\includegraphics[scale=1]{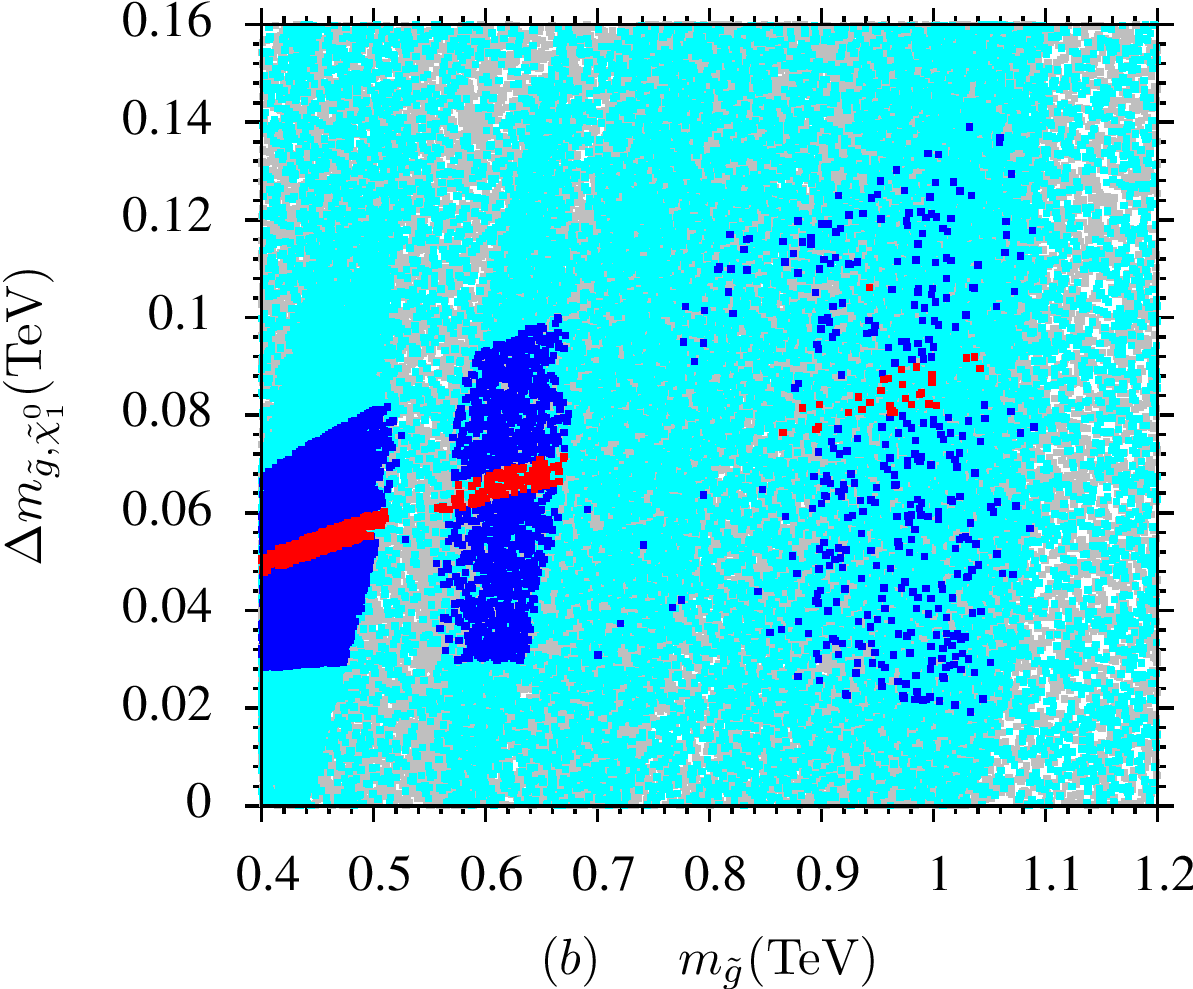}}
\subfigure{\includegraphics[scale=1]{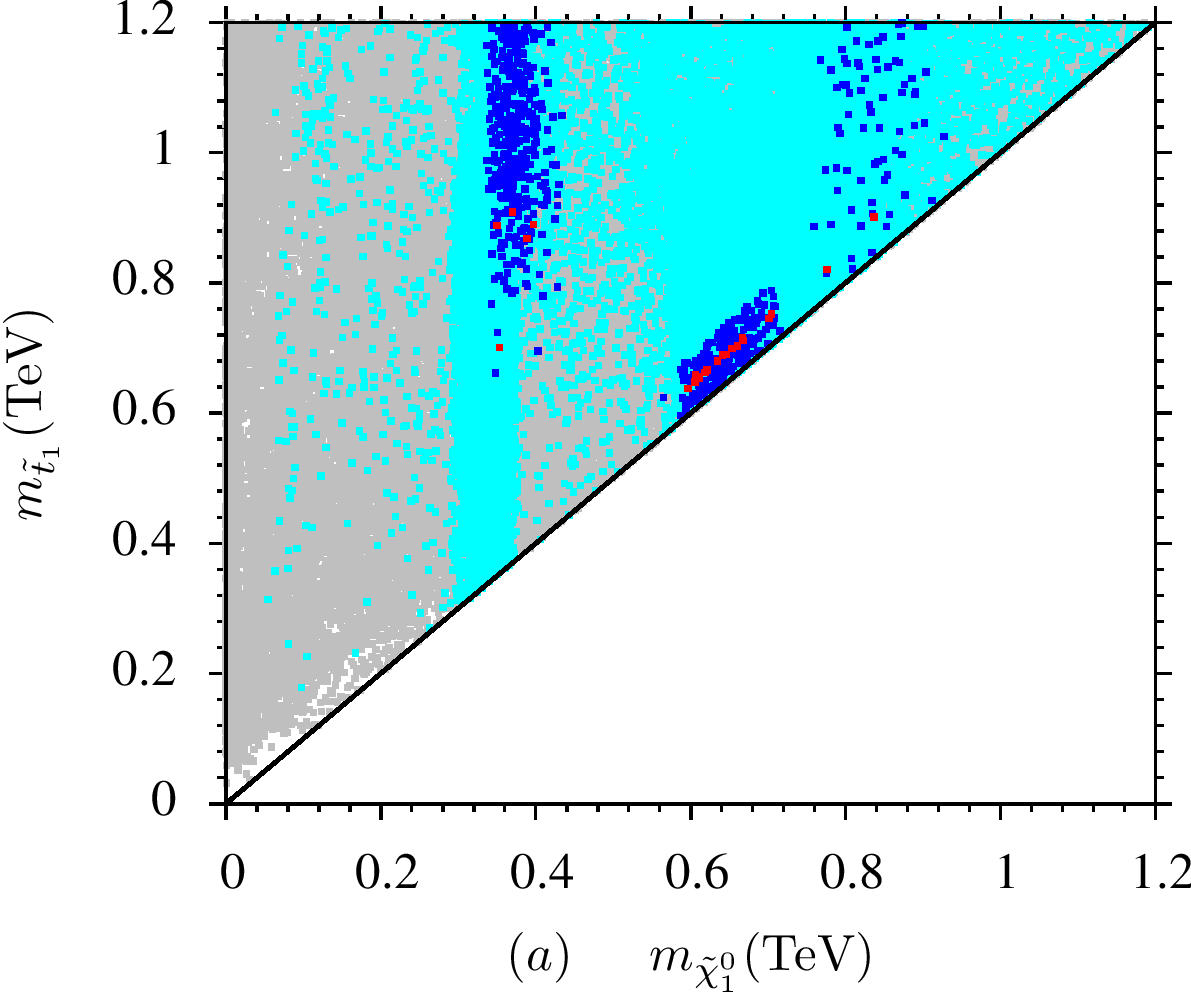}}
\subfigure{\includegraphics[scale=1]{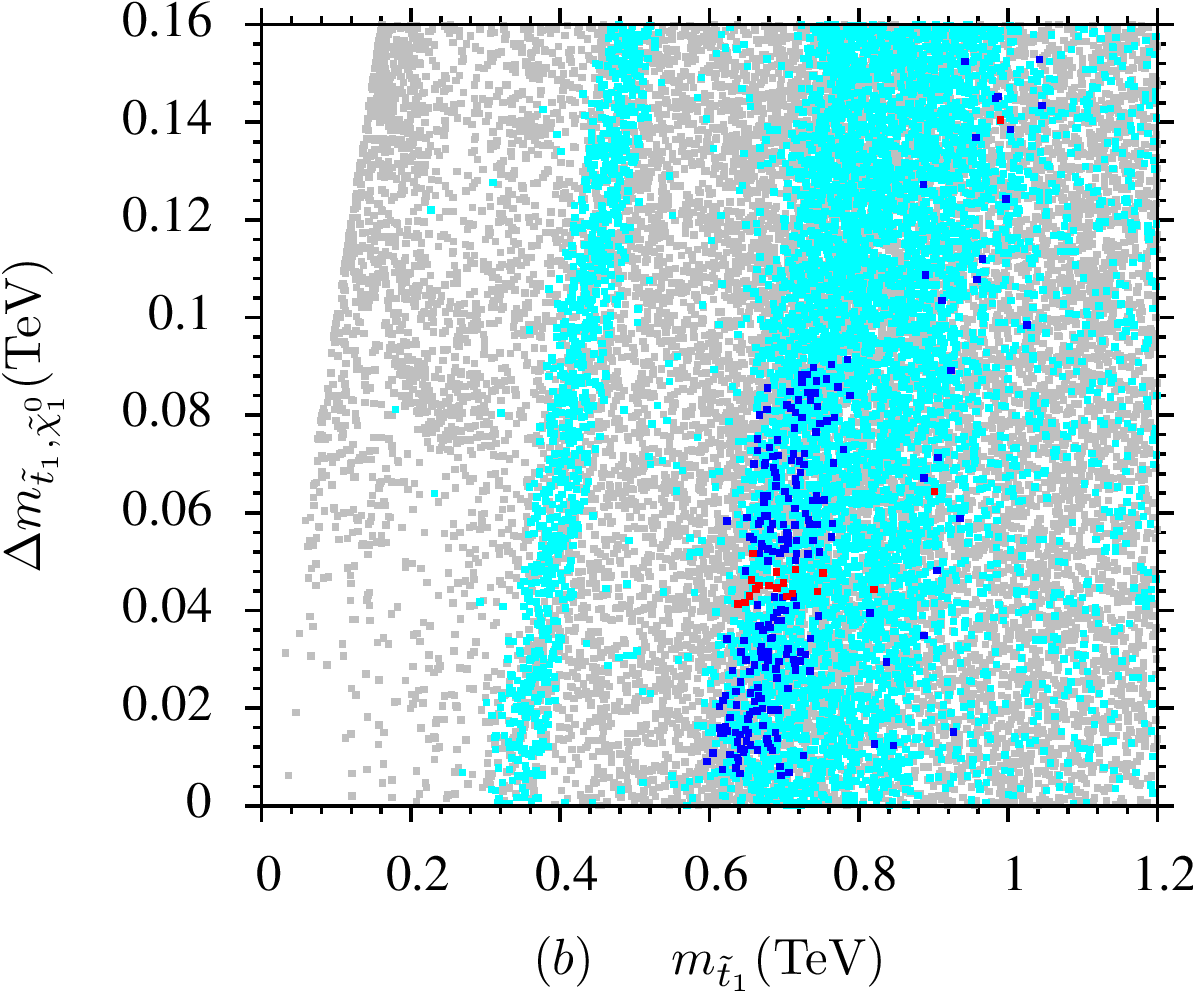}}
\caption{Plots in $m_{\tilde{g}}-m_{\tilde{\chi}_{1}^{0}}$, $\Delta m_{\tilde{g},\tilde{\chi}_{1}^{0}}-m_{\tilde{g}}$, $m_{\tilde{t}_{1}}-m_{\tilde{\chi}_{1}^{0}}$, $\Delta m_{\tilde{t}_{1},\tilde{\chi}_{1}^{0}}-m_{\tilde{t}}$ planes where $\Delta m_{\tilde{g},\tilde{\chi}_{1}^{0}}=m_{\tilde{g}}-m_{\tilde{\chi}_{1}^{0}}$ and $\Delta m_{\tilde{t}_{1},\tilde{\chi}_{1}^{0}}=m_{\tilde{t}_{1}}-m_{\tilde{\chi}_{1}^{0}}$. Grey points are consistent with REWSB and LSP neutralino. Aqua points satisfy mass bounds including bounds on Higgs mass and B-physics constraints. Blue points belong to a subset of aqua points and represent solutions with $\Omega h^{2} \leq 1$ and $R_{b\tau}\leq 1.1$. Red points form a subset and they are consistent with WMAP9 bound within $5\sigma$.}
\label{btau-nlsp}
\end{figure}
Figure \ref{btau-nlsp} displays plots in $m_{\tilde{g}}-m_{\tilde{\chi}_{1}^{0}}$, $\Delta m_{\tilde{g},\tilde{\chi}_{1}^{0}}-m_{\tilde{g}}$, $m_{\tilde{t}_{1}}-m_{\tilde{\chi}_{1}^{0}}$, $\Delta m_{\tilde{t}_{1},\tilde{\chi}_{1}^{0}}-m_{\tilde{t}}$ planes, where $\Delta m_{\tilde{g},\tilde{\chi}_{1}^{0}}=m_{\tilde{g}}-m_{\tilde{\chi}_{1}^{0}}$ and $\Delta m_{\tilde{t}_{1},\tilde{\chi}_{1}^{0}}=m_{\tilde{t}_{1}}-m_{\tilde{\chi}_{1}^{0}}$ . Grey points are consistent with REWSB and LSP neutralino. Aqua points satisfy various mass bounds including bounds on the Higgs mass and B-physics constraints. Blue points belong to a subset of aqua points and represent solutions with $\Omega h^{2} \leq 1$ and $R_{b\tau}\leq 1.1$. Red points form a subset and they are consistent with the WMAP9 bound within $5\sigma$. Let us discuss these 
graphs in some details. In $t-b-\tau$ and $b-\tau$ YU, the third generation squarks are relatively light compared to those of the first two families. As a consequence, the gluino decay may lead to top-rich or bottom-rich decay signals. In the coannihilation region where $\Delta m_{\tilde{g},\tilde{\chi}_{1}^{0}} \ll 2m_{t}$, there is no phase-space for on-shell top quarks. The gluino in this case decays into b-jets, $\tilde{g}\rightarrow b\bar{b}\tilde{\chi}_{1}^{0}$, 
which enables one to search for the NLSP gluino via multi-b jets, namely

\begin{equation}
pp \rightarrow \tilde g \tilde g \rightarrow b\bar b b \bar b + \cancel E_{T}.
\end{equation}
Such a scenario is favored for $\Delta m_{\tilde{g},\tilde{\chi}_{1}^{0}} \lesssim 100$ GeV. Note that the previous studies have ruled out a NLSP gluino 
with mass below 300 GeV \cite{Ajaib2}. 


We see from the top panels of Figure \ref{btau-nlsp} that the results for NLSP gluino with $b-\tau$ YU are similar to those obtained in the case of $t-b-\tau$ YU (as shown in next section). For a NLSP gluino mass of order a TeV or so the mass difference with the LSP dark matter neutralino should be at least 50 GeV in order to be consistent with the WMAP9 bound (within $5\sigma$) on dark matter relic abundance. In the region where the NLSP gluino is almost degenerate with the LSP neutralino ($\Delta m_{\tilde{g},\tilde{\chi}_{1}^{0}}\simeq 0$), the relic abundance of the latter is heavily reduced through coannihilation thus making it inconsistent with the WMAP9 bound. It can be seen that our results with $m_{\tilde g} \gtrsim 800$ GeV avoid the exclusion limits reported in \cite{ATLAS_nlsp_g,CMS_nlsp_g}. We also note that
according to recent studies \cite{Bhattacherjee:2013wna,Mukhopadhyay:2014dsa,Cohen:2013xda,Harigaya:2014dwa,Harigaya:2014pqa},
 our results can be readily tested at the LHC. It is indicated in \cite{Cheng:2014taa} that in certain scenarios an LSP with mass $\gtrsim$ 600 GeV
may evade the current LHC SUSY searches. 
 
The bottom panels indicate a distinct property of $b-\tau$ YU in $4-2-2$ not found in $t-b-\tau$ YU. We found in the $b-\tau$ case NLSP stop solution, with masses for the latter of order $\sim 600 - 900$ GeV. Note that NLSP stop solutions were previously found in the study of $b-\tau$ YU in SUSY $SU(5)$ in the mass range of $\sim 100-400$ GeV with $\tan\beta \gtrsim 20$ \cite{Baer:2012by}. Our results are in agreement with the results previously reported in \cite{Gogoladze:2011be} and \cite{Ellis:2014ipa}.

The bottom right panel shows that the mass difference between the NLSP stop and LSP neutralino should be at least 40 GeV to satisfy the WMAP9 dark matter abundance bound within $5\sigma$. The search for NLSP stop in such a scenario is challenging and has been implemented both at LEP and Tevatron \cite{Nakamura:2010zzi,jose,cdf}. The two-body stop decay into a top quark and neutralino or a bottom quark and chargino, and the three-body decay channels $\tilde t_1 \rightarrow W^{+}b{\tilde \chi_{1}^0}$,
$\tilde t_1 \rightarrow bl^{+}\nu$ are kinematically not allowed. The loop induced two-body decay of NLSP stop, $\tilde t_{1}\rightarrow c \chi_{1}^{0}$,
is generally considered to overwhelm the four-body channel $\tilde t_1 \rightarrow l^{+}\nu(q\bar q^{'})b{\tilde \chi_{1}^0}$ and tends to be the dominant NLSP stop decay mode
~\cite{kobayashi}. Results from searches for this decay channel using the Tevatron data have been reported by both the CDF and D0 experiments \cite{Aaltonen:2012tq,Abazov:2008rc}. Both model independent and model dependent studies of stop-neutralino coannihilation show that regions of the parameter space with stop-neutralino mass difference of $20\%$ are ruled out for $m_{\tilde{t}_{1}} \lesssim 140$ GeV \cite{Ajaib:2011hs,He}. Also, the first LHC searches for stop decaying into a charm quark and neutralino have recently been performed by the ATLAS collaboration \cite{ATLAS_NLSPstop} and stop masses up to 270 GeV have been excluded for LSP neutralino mass of about 200 GeV. On the other hand, $m_{\tilde t_1}$= 250 GeV with $\Delta m_{\tilde t_{1},\tilde \chi_1^{0}} \le$ 10 GeV has been ruled out by a recent CMS analysis \cite{CMS_NLSPstop}. The NLSP stop mass obtained from our analysis lies well beyond these exclusion limits, but we hope that the future LHC searches will probe it. We also find regions in the parameter space in which the stop and gluino masses are almost degenerate. 

\subsection{NLSP gluino from $t$-$b$-$\tau$ YU}

\begin{figure}[htp!]
\subfiguretopcaptrue

\subfigure{
\includegraphics[totalheight=5.5cm,width=7.0cm]{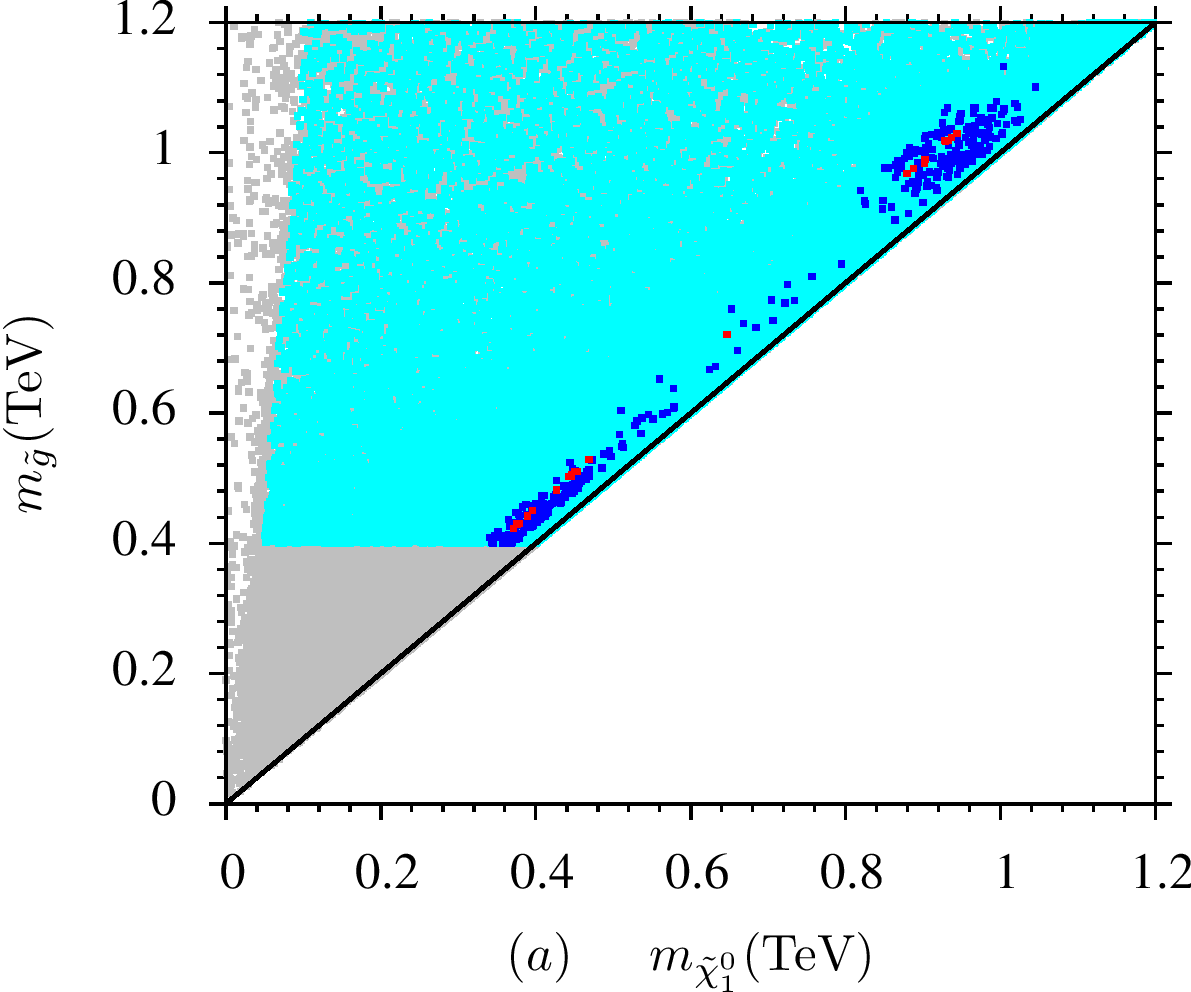}
}%
\subfigure{
\includegraphics[totalheight=5.5cm,width=7.0cm]{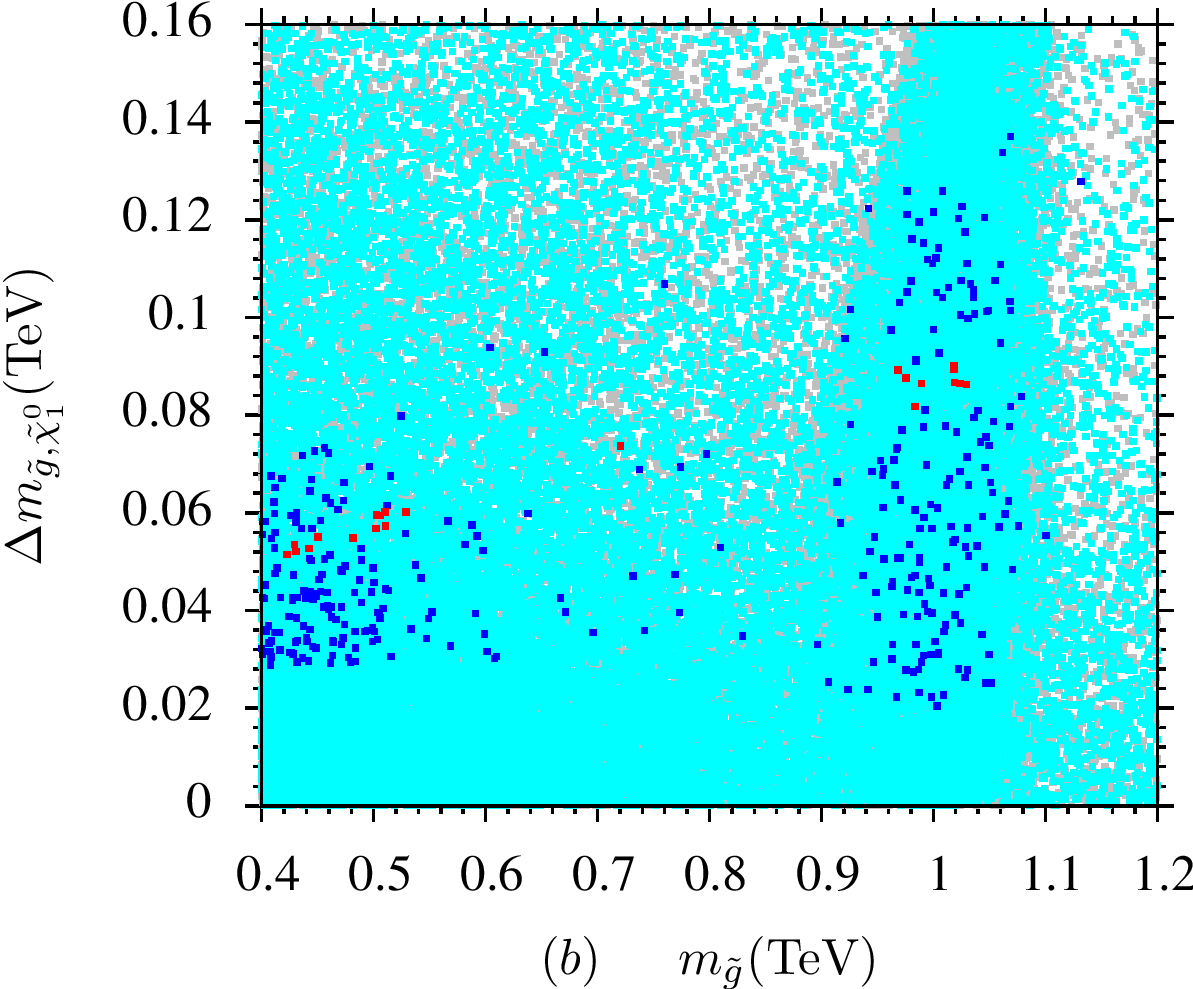}
}
\caption{Plots in $m_{\tilde{g}}-m_{\tilde{\chi}_{1}^{0}}$ and $m_{\tilde{g},\tilde{\chi}_{1}^{0}}$ planes. Color coding is the same as in Figure \ref{btau-nlsp}, except $R_{b\tau} \leq 1.1$ condition is replaced with $R_{tb\tau} \leq 1.1$.}
\label{nlsp_g}
\end{figure}

In this section, we revisit $t-b-\tau$ YU in $4-2-2$ to update the results taking account of the current experimental constraints, and then compare with the results obtained for $b-\tau$ YU. We quantify $t-b-\tau$ YU with $R_{tb\tau}$ defined in the same way as was done for $b-\tau$ YU:

\begin{equation}
\label{eq:Rtbtau}
R_{tb\tau}=\frac{ {\rm max}(y_{t},y_b,y_{\tau})} { {\rm min} (y_{t},y_b,y_{\tau})}.
\end{equation}

\vspace{0.8cm}

We summarize our results for NLSP gluino in Figure \ref{nlsp_g}. Color coding is the same as in Figure \ref{btau-nlsp}, except that the condition $R_{b\tau} \leq 1.1$ is replaced with $R_{tb\tau} \leq 1.1$. The left panel shows that NLSP gluino with $m_{\tilde{g}} \gtrsim 1$ TeV can be realized consistent with $R_{tb\tau} \leq 1.1$ and all the experimental constraints. 
The NLSP gluino solutions consistent with 10$\%$ or better $t-b-\tau$ YU posses more or less the same 
features as previously discussed for $b-\tau$ YU.

Finally we present five benchmark points in Table~\ref{table1} highlighting phenomenologically interesting
features of the Yukawa unified 4-2-2 model. All of these benchmark points satisfy the various constraints mentioned in 
Section ~\ref{sec:scan} and are compatible with Yukawa unification. Points 1-4 are the examples of 10$\%$ or better $b$-$\tau$ YU. Points 1 and 2 display NLSP gluino solutions with $BF(\tilde g \rightarrow b \bar b \chi_{1}^{0}) \approx$ 0.58. 
Point 3 represents an example where gluino is NLSP with $BF(\tilde g \rightarrow g \chi_{1}^{0}) \approx$ 0.83 and 
stop is NNLSP while point 4 depicts the opposite example with 
$BF(\tilde t_1 \rightarrow c \chi_{1}^{0})\approx$ 1.00. Point 5 is an example of $t-b-\tau$ YU and exhibits a 
relatively heavy NLSP gluino solution with $R_{tb\tau} \approx 1.09$ and $BF(\tilde g \rightarrow b \bar b \chi_{1}^{0})\approx$ 0.77. 
\begin{table}[hp!]
\centering
\begin{tabular}{|l|c|c|c|c|c|}
\hline
\hline
                 &  Point 1 & Point 2  &  Point 3  & Point 4 & Point 5   \\
\hline
$m_{16}$        &  19280   & 19460   & 11670    & 11940 & 19090      \\
$M_{1}$         &  1864.96 & 1762.92 & 1729.84   & 1700  & 1795      \\
$M_{2} $          & 2945    & 2790    & 2711   & 2660   & 2844     \\
$M_{3}$         &  244.9  & 222.3   & 258.1   & 260.1   & 221.4  \\
$A_{0}/m_{16}$  &  -2.691  & -2.685  & -2.631   & -2.632  & -2.52   \\
$\tan\beta$       & 37.8    & 37.7    &  38.9  &  39   & 50.7  \\
$m_{H_d}$         & 10890   & 11820   & 3246   &  3111  & 15640   \\
$m_{H_u}$          & 4931    & 4555    & 5175   & 5478  & 9938 \\
\hline
$m_h$                & {\color{red}124}   & {\color{red}125}    & {\color{red}124}   & {\color{red}123}  & {\color{red}126} \\
$m_H$                 & 11254 & 12176  & 3357  & 3058 & 3253   \\
$m_A$                 & 11181 & 12097  & 3336  & 3039  & 3232       \\
$m_{H^{\pm}}$         & 11255 & 12177  & 3359  & 3061 & 3255  \\
\hline
$m_{\tilde{\chi}^0_{1,2}}$
                &  {\color{red}952}, 2794 & {\color{red}903}, 2659 & {\color{red}850}, 2481  & {\color{red}837}, 2441 & {\color{red} 928}, 2728   \\
$m_{\tilde{\chi}^0_{3,4}}$
               & 22197, 22197 & 22407, 22407&12855, 12855 &13101, 13101 & 19151, 19151    \\

$m_{\tilde{\chi}^{\pm}_{1,2}}$
               &  2891, 22246 & 2674, 22453& 2497, 12844  &2457, 13090  & 2735, 19151   \\
\hline
$m_{\tilde{g}}$ &  {\color{red}1041}   & {\color{red}988}  & {\color{red}933}  & {\color{red}943} & {\color{red}1018}           \\
$m_{ \tilde{u}_{L,R}}$
               & 19347, 19216& 19520, 19374& 11757, 11702  &12019, 11977  & 19174, 18989     \\
$m_{\tilde{t}_{1,2}}$
               & 3107, 8349& 3426, 8509 & {\color{red}991}, 5176 & {\color{red}901}, 5263  & 5692, 7556     \\
\hline $m_{ \tilde{d}_{L,R}}$
                &19347, 19334 & 19520, 19522 &11758, 11666  &12019, 11933 & 19174, 19174        \\
$m_{\tilde{b}_{1,2}}$
                 &8405, 11088 & 8560, 11227 &5205, 6881  &5293, 7047 &7414, 19098    \\
\hline
$m_{\tilde{\nu}_{1}}$
              &  19332  &19487 & 11824 &12091 & 19104       \\
$m_{\tilde{\nu}_{3}}$   & 15837& 15962  & 9848 &10076 & 14208    \\
\hline
$m_{ \tilde{e}_{L,R}}$
              & 19319, 19394 & 19474, 19600 & 11815 &12081, 11906 & 19098, 19267    \\
$m_{\tilde{\tau}_{1,2}}$
              &  11668, 15838 & 11841, 15968 & 11645  &7367, 10045 & 6806, 14137    \\
\hline

$\sigma_{SI}({\rm pb})$
             &  2.49$\times 10^{-14}$& 2.73$\times 10^{-14}$ & 7.44$\times 10^{-15}$ & 2.77$\times 10^{-14}$ & $1.67\times 10^{-14}$ \\

$\sigma_{SD}({\rm pb})$
             &  2.52$\times 10^{-14}$ &2.82$\times 10^{-14}$ &1.1$\times 10^{-16}$   &6.09$\times 10^{-17}$  & $8.5 \times 10^{-15}$\\
$\Omega_{CDM}h^{2}$
            &  0.116 & 0.102 &0.112  & 0.122  & 0.124\\
\hline
$R_{b\tau}$, $R_{tb\tau}$         & 1.09 & 1.09  &1.09  & 1.09 & 1.09 \\
\hline
\hline
\end{tabular}
\caption{
Sparticle and Higgs masses (in ${\rm GeV}$ units). Fundamental parameters are specified at $M_{{\rm GUT}}$. All of these benchmark points satisfy
the various constraints mentioned in Section~\ref{sec:scan} and are compatible with Yukawa
unification. Points 1-4 are examples of 10$\%$ or better $b$-$\tau$ YU.
Points 1 and 2 display NLSP gluino solutions. Point 3 represents an example where gluino is NLSP and
stop is NNLSP with a small mass difference while point 4 depicts the opposite example. Point 5 exhibits a relatively heavy
NLSP gluino solution with $R_{tb\tau} = 1.09$.}
\label{table1}
\end{table}

\newpage

\section{Conclusion}
\label{sec:conclude}

We have explored $b-\tau$ and $t-b-\tau$ YU in supersymmetric $SU(4)_c \times SU(2)_L \times SU(2)_R$ (4-2-2) models with the MSSM parameter $\mu >$ 0.  Our results extend earlier discussions of 4-2-2 models and can be tested at LHC 14. We show that NLSP gluino masses of order 1 TeV are compatible with
 $b-\tau$ or $t-b-\tau$ Yukawa unification,while NLSP gluino masses of order 300 GeV or lower have now been excluded. We also display solutions in 
$b-\tau$ Yukawa unified models with NLSP stop masses $m_{\tilde t_1} \gtrsim$ 600 GeV. In such cases, the mass difference between NLSP stop and LSP neutralino allows only the decays $\tilde{t}_{1}\rightarrow c\tilde{\chi}_{1}^{0}$. This type of decay has recently been studied by the ATLAS and CMS collaborations, and our conclusions are consistent with their results. We also identify an interesting region of parameter space where NLSP and NNLSP masses are almost degenerate. In this region either the stop or gluino is the NLSP. We also revisit supersymmetric $4-2-2$ model with $t-b-\tau$ YU which
yields neutralino-gluino coannihilation solutions. This is the only channel compatible with the observed dark matter relic abundance. We 
find that NLSP gluino solutions in such a case have the same features as NSLP gluino in the $b-\tau$ YU scenario. We present five benchmark points as representatives of our solutions that may be tested in future LHC experiments.

\section*{Acknowledgement}
We would like to thank Bin He, Jinmian Li, Tianjun Li and Jacob Wacker for useful discussions. This work is supported by the DOE Grant No. DE-FG02-91ER40626 (S.R and Q.S).
This work used the Extreme Science and Engineering Discovery Environment (XSEDE), 
which is supported by the National Science
Foundation grant number OCI-1053575


\end{document}